\documentclass{aa}
\usepackage{graphicx}
\usepackage{txfonts}
\usepackage{psfrag}
\usepackage{longtable}
\usepackage{graphics}
\usepackage{amssymb}

\begin{document}

\title{Detection of optical linear polarization in the SN~2006aj/XRF~060218
  non-spherical  expansion \thanks{Based  on observations  made  with the
    Nordic Optical Telescope,  operated on the island of  La Palma, in
    the  Spanish  Observatorio  del  Roque  de los  Muchachos  of  the
    Instituto  de Astrof\'{\i}sica  de Canarias,  and  on observations
    done at the Centro Astron\'omico Hispano Alem\'an (CAHA) at
    Calar  alto, operated  jointly  by the  Max-Planck Institut  f\"ur
    Astronomie and the Instituto de Astrof\'{\i}sica de Andaluc\'{\i}a
    (CSIC).}}

\author{J.~Gorosabel\inst{1}
   \and V.~Larionov\inst{2}
   \and A.J.~Castro-Tirado\inst{1}
   \and S.~Guziy\inst{1,3}
   \and L.~Larionova\inst{2}
   \and A. Del Olmo\inst{1}  
   \and M.A.~Mart\'{\i}nez\inst{1}  
   \and J. Cepa\inst{4}
   \and B. Cedr\'es\inst{4}
   \and A. de Ugarte Postigo\inst{1}
   \and M. Jel\'\i nek\inst{1}
   \and O. Bogdanov\inst{3}
   \and A. LLorente\inst{5}
}

\offprints{ \hbox{J. Gorosabel, e-mail:{\tt jgu@iaa.es}}}

\institute{Instituto de Astrof\'{\i}sica de Andaluc\'{\i}a (IAA-CSIC),
   Apartado de Correos, 3.004, E-18.080 Granada, Spain.
  \and
   Astronomical Institute of  St. Petersburg University, Petrodvorets,
   Universitetski pr. 28, 198504, Russia.
  \and
   Nikolaev State University, Nikolskaja 24, Nikolaev, 54030, Ukraine.
  \and
   Instituto de Astrof\'{\i}sica de Canarias, La Laguna, Tenerife, E-38200 Canary Islands, Spain.
  \and
   XMM-Newton Science Operations Centre, European Space Agency, 
   Villafranca del Castillo, PO Box 50727, 28080 Madrid, Spain.}

\date{Received / Accepted }

\abstract {}  {We have performed optical  polarimetric observations of
  the SN~2006aj  associated to the  X-ray flash (XRF) of  February 18,
  2006, \object{XRF~060218} that  provide information on its expansion
  geometry.}  {The  data were acquired  in the $R$-band with  the 0.7m
  telescope of Crimea,  2.5m Nordic Optical Telescope and  the 2.2m of
  Calar  Alto.}   {We  report  the detection  of  linear  polarization
  between 3  and 39  days after the  gamma-ray event  ($t-t_0$).  This
  represents the  first polarization detection of a  Ic supernova (SN)
  associated  to  an  XRF.   Our  data  exhibit  a  degree  of  linear
  polarization  ($P$)  around $P\sim4\%$  at  $t-t_0  \sim 3-5$  days,
  followed by  a constant  polarization phase with  $P \sim  1.4\%$ at
  $13.7 \lesssim t-t_0 \lesssim 39$ days.  Our data suggest a decay in
  $P$,  and  more  interestingly,  show a  position  angle  ($\theta$)
  rotation of $\sim 100^{\circ}$ comparing data taken before and after
  the   $R$-band  lightcurve   peak.}    {The  reported   polarization
  measurements can be  explained by the evolution of  an asymmetric SN
  expansion.  We discuss on several ingredients that could account for
  the observed $\theta$ rotation. }

\keywords{Gamma rays: bursts -- Supernovae: general -- Techniques: polarimetric}

\maketitle

\section{Introduction}

The {\it  Swift} satellite detected a long-soft  gamma-ray burst (GRB)
at $t_0$ = Feb.  18.149 2006 UT, (Cusumano et al.  2006), showing both
a low redshift ($z=0.033$, Mirabal  \& Halpern 2006) and a peak energy
(Campana et al.  2006),  typical of X-ray flashes (\object{XRF~060218}
hereafter).    Further  observations  associated   it  to   a  Ic-type
supernova,   \object{SN~2006aj}   (Masetti   et  al.    \cite{Mase06};
Soderberg et  al.  \cite{Sode06};  Pian et al.   \cite{Pian06}).  This
finding  strengthened  even   more  the  already  solid  long-duration
GRB/Supernova   link   (Hjorth   et   al  \cite{Hjor03};   Stanek   et
al. \cite{Stan03}).

Since 1999 (Covino et  al.  \cite{Covi99}) several GRB afterglows have
shown optical polarization at a level of $\sim$1-3\% (see Gorosabel et
al.  \cite{Goro04} and references therein).  Optical supernovae (SNe),
including typical  Ic-types not related to  GRBs, do not  tend to show
optical polarization  above $\sim1$\%.  However,  some counterexamples
exist (Leonard  et al.  \cite{Leon06}).  The  very few polarimetric
  studies of  Ic SNe related to GRBs  do not allow yet  to infer clear
  global polarimetric differences with  normal Ic SNe.  However, as we
  report in  this Letter, some  normal Ic SNe  like \object{SN~2002ap}
  (Mazzali  et  al.    \cite{Mazz02})  share  some  similarities  with
  \object{SN~2006aj};  polarization levels  above $\sim1$\%  and, more
  interestingly,  long-term   rotations  in  the   polarization  angle
  (Kawabata et al.  \cite{Kawa02}).

\vspace{-0.4cm}

 Most of  the SN intrinsic  polarization detections are  attributed to
 the break  of the symmetry  around line of  sight, due to  a Thompson
 photon scattering  through an aspherical  SN expansion (Kasen  et al.
 \cite{Kase03}).   This polarization  is  expected to  decay as  $\sim
 t^{-2}$, as the  atmosphere expands and the optical  depth (and hence
 the  polarization) falls  (Leonard  et al.   \cite{Leon06}). In  most
 cases the afterglow is so  strong that overshines the SN polarization
 component   (for    instance   in   GRB~030329;    Greiner   et   al.
 \cite{Grei03}).   In this  Letter  we report  the first  polarization
 detection to date  of a SN associated to an XRF,  not coming from the
 GRB ejecta forward shock emission (Fan et al. \cite{Fan06}; Soderberg
 et   al.   \cite{Sode06}).   Sect.~\ref{Observations}   reports  the
   observations      carried      out     for      \object{SN~2006aj}.
   Sect.~\ref{results} presents our results and Sect.~\ref{Discussion}
   discusses     the     implications     of     our     measurements.
   Sect.~\ref{Conclusions} summarises the final conclusions.

\section{Observations and data analysis}
\label{Observations}

Table~\ref{log}  displays  the  log  of  our  observations.   All  the
measurements reported in this paper  were carried out in the $R$-band.
The  observations were  performed, chronologically  ordered,  with the
0.7m telescope  of Crimea (AZT-8),  the 2.5m Nordic  Optical Telescope
(NOT) and the 2.2m of Calar Alto (CAHA).

The AZT-8 observations were done with two Savart plates; swapping them
the observer  can obtain either the  Stokes $Q$ (two  images are split
diagonally) or  the $U$ parameter  (split horizontally). The  field of
view (FoV)  covered by the  AZT-8 (+ST-7 CCD) is  $8.1^{\prime} \times
5.4^{\prime}$ and the pixel scale is 1.3$^{\prime \prime}$/pix.  Since
the  GRB  field is  substantially  crowded  in  the AZT-8  images,  we
performed PSF photometry (Stetson \cite{Stet87}).

\begin{figure}[t]
\begin{center}
\resizebox{\hsize}{4.0cm}{\includegraphics[bb=55 372 550 685]{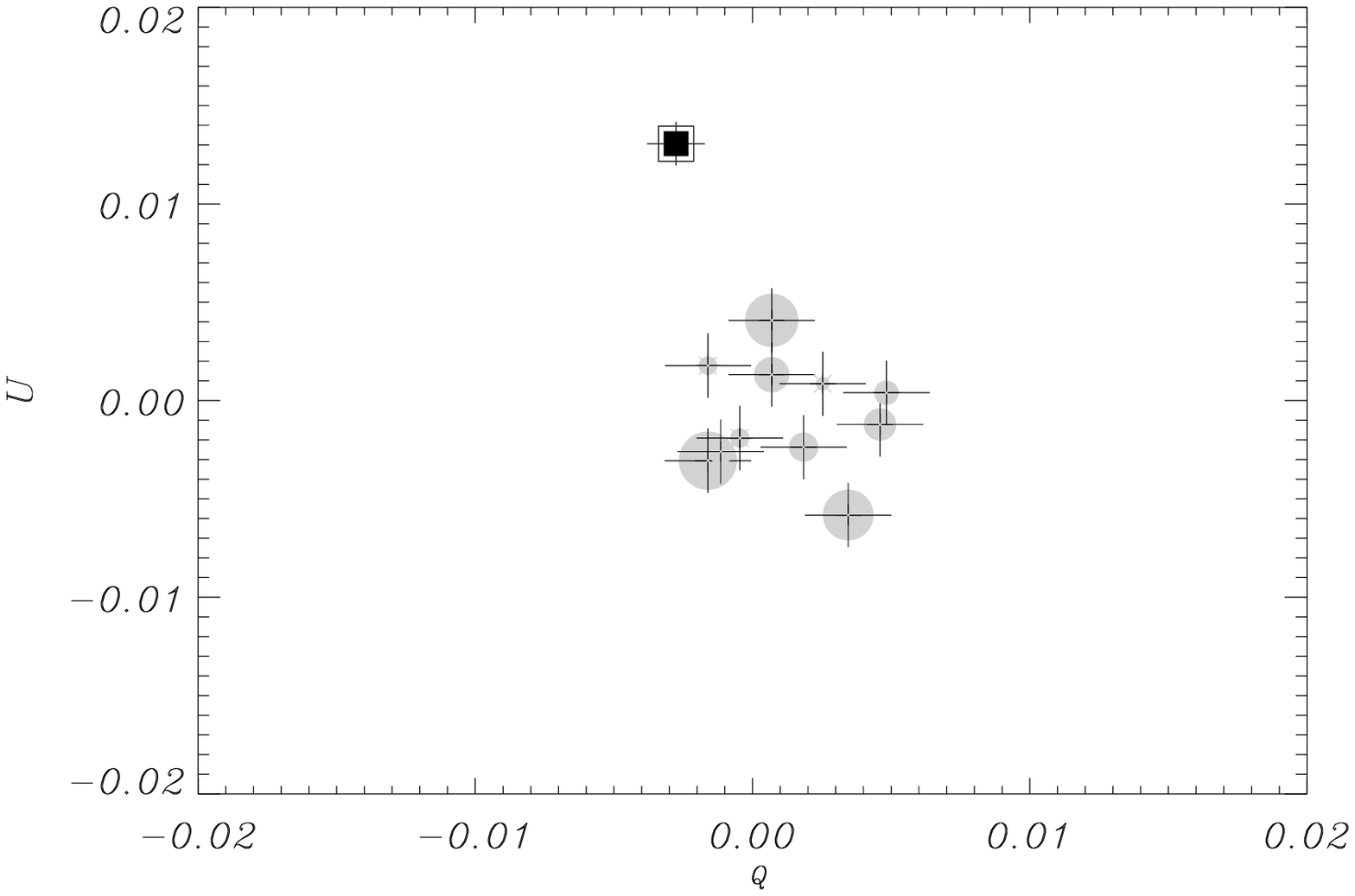}}
\resizebox{\hsize}{4.0cm}{\includegraphics[bb=55 372 550 695]{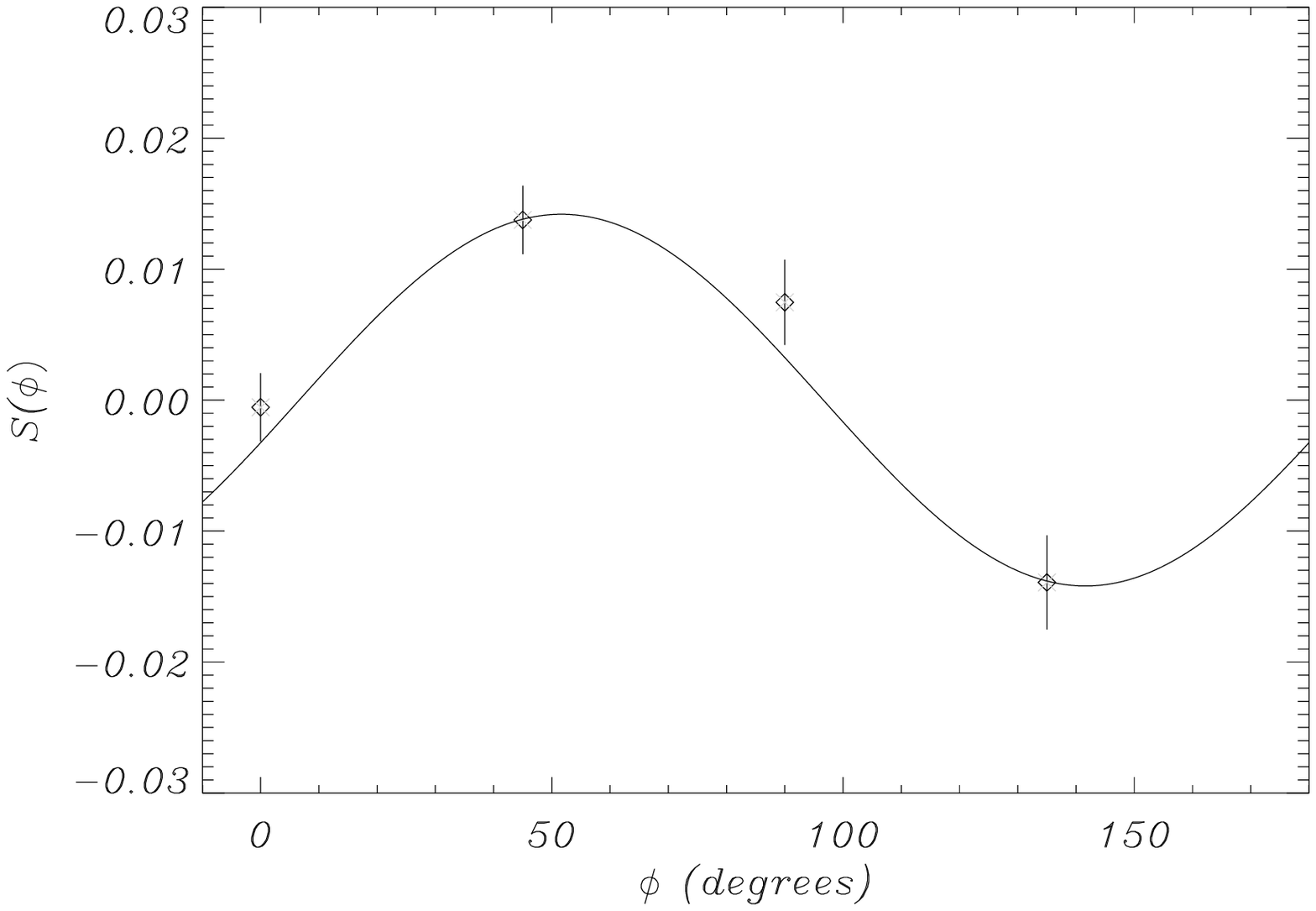}}
\caption{\label{fig1} {\em Upper  panel:} The Q, U plane  of the stars
  in the  field (circles) of \object{XRF~060218}  (square) imaged from
  CAHA on  March 8.845  UT ($t-t_0$  = 18.7 days).   The size  of each
  circle is proportional to the  distance of the corresponding star to
  the center of the CCD.  This  is used to check that the instrumental
  polarization,  if present, does  not vary  severely across  the FoV.
  {\em Lower  panel:} The satisfactory  fit of the  $S(\phi)$ function
  ($\chi^2/d.o.f=1.38$).   As shown  the \object{XRF~060218/SN~2006aj}
  is clearly polarized.  An aperture radius = 1 FWHM was assumed.}
\end{center}
\end{figure}

The  NOT observations  were based  on ALFOSC  equipped with  the FAPOL
unit.  In order  to obtain $Q$ and $U$ with ALFOSC,  the GRB field was
imaged through a calcite and a  1/2 wave plate. Four images of the GRB
field  were acquired,  rotating  the 1/2  wave  plate at  $0^{\circ}$,
22.5$^{\circ}$, 45.0$^{\circ}$ and 67.5$^{\circ}$.  The calcite plates
of FAPOL produce  a vignetted field of about  140$^{\prime \prime}$ in
diameter with  a pixel scale of 0.19$^{\prime  \prime}$/pix.  The CAHA
observations were  carried with CAFOS. The CAFOS  polarization unit is
similar to FAPOL  (is based on a Wollaston prism  instead of a calcite
plate), but uses  a strip mask on the focal  plane to avoid accidental
overlapping on  the CCD.   The total  FoV in CAFOS  is composed  by 14
strips  of $9^{\prime}\times18^{\prime\prime}$ each,  containing stars
enough  for  a  satisfactory  interstellar medium  (ISM)  polarization
correction. For  the NOT  and CAHA data  the determination of  $P$ and
$\theta$  was carried out  fitting the  $S(\theta)$ function  with the
corresponding   ISM   normalization   factor  (di   Serego   Alighieri
\cite{dise97};  see  Fig.~\ref{fig1}).   For  both  data sets  aperture
photometry was performed,  ranging the aperture radii from  1 pixel to
2.5 times the FWHM (Full  Width at Half Maximum).  After checking that
the  magnitudes  were independent  of  the  apertures  used, the  ones
yielding minimum errors were adopted.

The Galactic ISM  polarization correction was done using  9 (13) field
stars on the AZT-8 (CAHA) images.   Due to the reduced ALFOSC FoV only
one unsaturated bright field star was available in the NOT data.  Thus
for the NOT images zero (G191B2B) and high (HD25443) polarization
standards allowed us to infer the \object{SN~2006aj} Stokes parameters,
and for  completeness, also the  ones of the unique  unsaturated field
star   (which  agreed   with  the   ISM  law   by  Serkowski   et  al.
\cite{Serk75}).

\section{Results}
\label{results}

\begin{figure}[t]
\begin{center}
\resizebox{6.6cm}{5.4cm}{\includegraphics[bb=80 160 602 590]{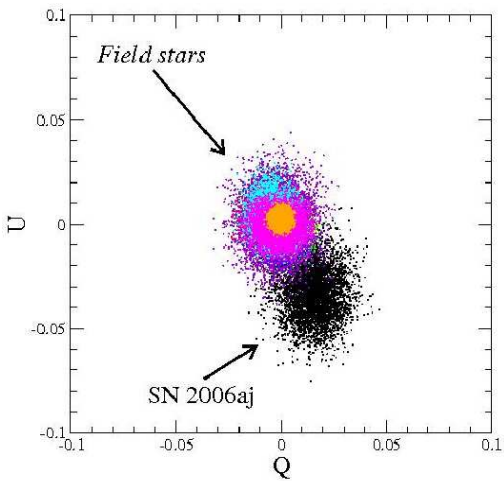}}
\caption{\label{fig2} The Stokes plane  of 9 field stars and SN~2006aj
  on  the combined  AZT-8 image  when the  Galactic ISM  correction is
  applied.  The clouds of dots were created altering with a Montecarlo
  method  the fluxes  of the  objects according  to  their photometric
  errors.  The cloud representing SN~2006aj is shifted with respect to
  the origin defined  by the field stars, assumed  to be intrinsically
  unpolarized in average.}
\end{center}
\end{figure}

Table~\ref{log} shows the position  angle ($\theta$) and the degree of
linear polarization ($P$) of the afterglow once they were corrected by
the  ISM polarization  and also  by the  fact that  $P$ is  a positive
quantity  (multiplying  $P$  by  $\sqrt{1-(\sigma_P/P)^{2}}$;  see  di
Serego  Alighieri  \cite{dise97}).   The  AZT-8 results  displayed  in
Table~\ref{log}  supersede the ones  reported previously  (Larionov \&
Larionova \cite{Lari06}). The four  AZT-8 data sets yield $P$ marginal
detections,  reaching a $3\sigma$  confidence level  at $t-t_0\sim3.6$
days.  The apparently large $P$  variations (from $P$=2.43 to 4.53) in
the first  four nights are  within $P$ errors,  so we they  are likely
statistical fluctuations.   However, we note that for  the four epochs
\object{SN~2006aj} is  systematically placed at the  same Stokes plane
area, once the ISM correction  is included.  In order to reinforce the
$P$ detection, the  four AZT-8 observing epochs were  co-added and PSF
photometry performed.   The fifth  line of Table~\ref{log},  shows the
result when the  four AZT-8 epochs are combined.   The estimate of the
$P$ and  $\theta$ errors in  the AZT-8 images were  obtained modelling
with   a   Montecarlo   method   the  flux   error   distribution   of
\object{SN~2006aj} and the 9 field  stars used for the ISM correction.
Fig.~\ref{fig2} shows  the Stokes  plane when all  the AZT-8  data are
combined.   The  cloud  of  dots representing  the  SN~2006aj  ($Q,U$)
distribution is  clearly off  the center defined  by the  field stars.
The  $P$ value inferred  from the  combined image  is $3.40\pm1.09\%$,
detected     at     a     3.1$\sigma$     level     (including     the
$\sqrt{1-(\sigma_P/P)^{2}}$ term).

\begin{table*}
\begin{center}
    \caption{\label{log} Log of the  observations. For the NOT  and CAHA the
      observation are  cycles of 4  images, ($4 \times  T_{exp}$) with
      the plate  rotator at  0, 22.5, 45,  67.5 degrees.  The $R$-band
      magnitudes  are calibrated using the  standard stars
      given by Hicken et al. (\cite{Hick06}).}
\begin{tabular}{ccccccc}
   \hline 
 Days from GRB & Telescope    &Exposure Time& Seeing & Magnitude &$P$&$\theta$\\ 
   $t-t_0$  & (+Instrument)&  seconds    &  ``    & $R$-band  &\%&   degrees\\
   \hline
 2.569954 - 2.597558 &0.7m AZT-8(+ST-7)&$2 \times (10\times 90)$& 3.0 &$17.99\pm 0.03$&$2.61\pm1.80$&$157.82\pm 16.47$\\
 3.579178 - 3.600764 &0.7m AZT-8(+ST-7)&$2 \times (10\times 90)$& 3.8 &$17.88\pm 0.02$&$4.53\pm1.52$&$143.64\pm  7.18$\\ 
 4.555046 - 4.599144 &0.7m AZT-8(+ST-7)&$2 \times (20\times 90)$& 3.3 &$17.68\pm 0.02$&$4.06\pm1.63$&$143.86\pm  7.98$\\
 5.565510 - 5.612326 &0.7m AZT-8(+ST-7)&$2 \times (20\times 90)$& 4.2 &$17.52\pm 0.02$&$2.43\pm0.89$&$150.09\pm 10.02$\\
 2.569954 - 5.612326 &0.7m AZT-8(+ST-7)&$2 \times (60\times 90)$& 3.5 &$17.71\pm 0.02$&$3.40\pm1.09$&$147.56\pm  7.06^{\star}$\\
   \hline
13.721516 - 13.755185 &2.5m NOT(+ALFOSC)&$4 \times 700$& 1.0 &$17.19 \pm 0.04$&$1.44\pm0.14$&$49.58\pm 2.73$\\ 
18.673955 - 18.719135 &2.2m CAHA(+CAFOS)&$4 \times 900$& 1.6 &$17.38 \pm 0.03$&$1.40\pm0.21$&$51.62\pm 4.11$\\
38.679248 - 38.707514 &2.2m CAHA(+CAFOS)&$4 \times 400$& 2.0 &$18.44 \pm 0.03$&$<3.9^{\dagger}$&$--$\\ 
   \hline
   \multicolumn{7}{l}{$\star$ Determined combining the images of the above four observing epochs. \hspace{1.0cm} $\dagger$ 3~$\sigma$ upper limit.}\\
   \hline
\end{tabular}
\end{center}
\end{table*}

Both the NOT  and first CAHA data show  conclusive $P$ detections with
confidence levels  of $10.3\sigma$ and  $6.7\sigma$ respectively.  The
$\theta$  values of  both  epochs  agree with  a  stable angle  around
50$^{\circ}$,    but    interestingly    it   appears    rotated    by
$\sim100^{\circ}$ with  respect to the AZT-8 images.   Our data points
reject a  stationary $P$  value at a  3$\sigma$ level, so  we conclude
that $P$  likely fades at 5.6  $<t-t_0<$ 13.7 days.  Such  a $P$ decay
resembles  the  late evolution  of  the asymmetric  \object{SN~2004dj}
(Leonard et  al.  \cite{Leon06}), which  showed an abrupt $P$  peak at
$t-t_0\sim90$ days  due to the  sharp appearance of the  inner ejecta,
followed by a late $\sim t^{-2}$ decay.

 We  point  out that  the  result of  our  $R$-band  polarimetry is  a
 weighted  mean  of  the  continuum  and  the  spectral  features  (in
 principle each one with  a different polarization) falling inside the
 wavelength range  of the  filter transmittance (centred  at 6500\AA).
 With  no spectro-polarimetric  data it  is not  possible a  priory to
 quantify  if  the measured  polarization  evolution  is dominated  by
 spectrally  localized  large   polarization  fluctuations.   We  note
 however that the SN~2006aj  spectra around 6500\AA~ are smooth.  Thus
 the $\theta$  rotation measured at $5.5  <t-t_0 < 13.7$  days, is not
 easily explainable  by colour evolution, which  occurs at wavelengths
 bluer   than    5000\AA~   in   that   epoch    range   (Mazzali   et
 al. \cite{Mazz06}).
  
\section{Discussion}
\label{Discussion}

\begin{figure}[t]
   \resizebox{\hsize}{6.5cm}{\includegraphics[bb=30 60 709 595]{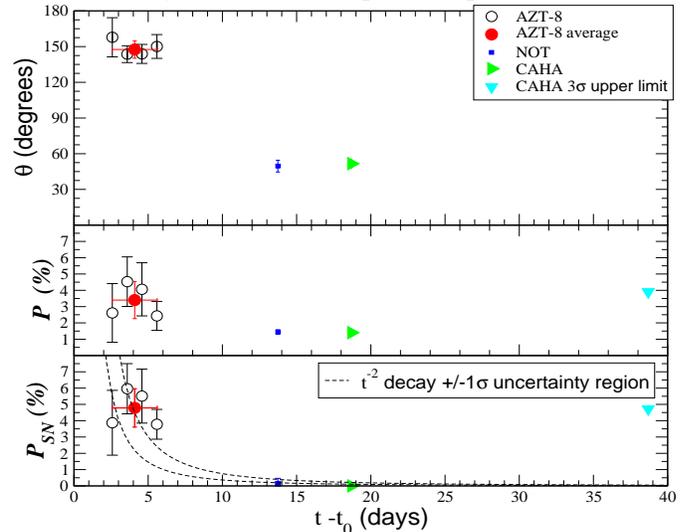}}
\caption{\label{fig3} Our polarization  results, once the the Galactic
  ISM correction was considered.  {\em Upper panel:} As shown $\theta$
  exhibited a  rotation of $\sim100^{\circ}$  at $5.6 < t-t_0  < 13.7$
  days.   {\em Middle  panel:} $P$  shows  a likely  decay reaching  a
  constant  level at  a $\sim  1.4\%$ level.   {\em Lower  panel:} The
  intrinsic $P_{SN}$ evolution assuming that our last $P$ detection is
  entirely originated  by the  host ISM.  The  dashed curves  show the
  $\pm 1\sigma$ uncertainty  region of the $t^{-2}$ decay,  as seen in
  \object{SN~2004dj}.  }
\end{figure}

The most natural framework to explain our polarization measurements is
a non  spherical SN expansion,  consistent with the  scenario proposed
for   \object{XRF~060218}    by   recent   studies    (Pian   et   al.
\cite{Pian06}).   It  is  interesting   to  note  that  other  Ic  SNe
exhibiting non-spherical expansions  have been linked to long-duration
GRBs (H\"oflich et al.   \cite{Hofl99}).  Furthermore there is growing
evidence that an important fraction of Ic SNe are triggered by bipolar
ejections  (Granot  \&  Ramirez-Ruiz  \cite{Gran04}).   It  is  widely
accepted that non-zero SN polarization measurements demand some degree
of expansion asphericity.  If the  projection of the SN along the line
of sight is  elliptical, a non canceled linear  polarization should be
present parallel to the minor  axis (Kasen et al.  \cite{Kase03}).  An
ellipsoidal expansion is fairly  consistent with the measured $P$ (see
$t^{-2}$  curve of  Fig.~\ref{fig3}).  However,  a  single ellipsoidal
model can not account for the measured $\theta$ rotation.

The  decay  in  $P$ seems  to  be  accompanied  with  a decay  of  the
photospheric  expansion velocity  (notice the  similarity  between our
Fig.~\ref{fig3} lower panel and Fig.~2 of Pian et al.  \cite{Pian06}).
This possible correlation between the photospheric expansion speed and
the polarization,  might imply  the existence of  a fast ($v  > 20000$
km~s$^{-1}$,  Pian  et al.   \cite{Pian06})  highly asymmetric  ejecta
responsible of the high polarization measured $2.6-5.6$ days after the
GRB.   This rapid  asymmetric ejecta  might be  aligned along  the GRB
jet. Several days later a slower ($v < 15000$ km~s$^{-1}$, Pian et al.
\cite{Pian06}) SN  bulk ejecta  would circularice the  geometry, which
would  explain the lower  $P$ of  our CAHA  and NOT  measurements.  We
could speculate that  this slower ejecta might be  in a toroidal-shape
equatorial expansion,  so the geometric  transition from the  rapid to
slow  ejecta   might  naturally  explain  the   detected  rotation  of
$\sim100^{\circ}$.   These  results  partially resemble  the  jet-like
geometry  proposed for \object{SN~2002ap},  which exhibited  lower $P$
values than  \object{SN~2006aj}, but with a  $\theta$ rotation similar
to  it ($\sim70^{\circ}$;  Kawabata et  al.  \cite{Kawa02}).  However,
other ingredients  can also  induce a rotation  in $\theta$:  $i)$ the
host  galaxy ISM,  $ii)$ a  non radial  distribution of  $^{56}$Ni and
$iii)$  a dusty  circumstellar  medium (CSM).   These  effects can  be
combined producing a complex picture.

  As  discussed  by Klose  et  al.   (\cite{Klos04}) the  polarization
  measured in afterglows  are sensitive to the host  galaxy ISM, so it
  might well affect our observations.  Thus, the $\theta$ stability at
  $t-t_0>13.7$ days could be the result of the $P$ induced by the host
  ISM,  so the  \object{SN~2006aj}  intrinsic polarization  ($P_{SN}$)
  would dominate  just before  our NOT data  point.  According  to the
  Serkowski's law  $E(B-V) \geq  0.15$ would be  required in  the host
  line  of sight  to account  for the  $P \sim  1.4 \%$  NOT  and CAHA
  measurements.  This $E(B-V)$ value  agrees fairly well with the host
  extinction estimated  by Modjaz et  al.  (\cite{Modj06}; $E(B-V)\sim
  0.05-0.11$), Sollerman et al.  (\cite{Sole06}; $E(B-V)\sim0.1-0.3$),
  and Campana et al.   (\cite{Camp06}; $E(B-V)=0.2$).  We note however
  the   lower   estimate   by   Guenther   et   al.    (\cite{Guen06};
  $E(B-V)=0.042$\footnote{See Sollerman  et al. (\cite{Sole06})  for a
    discussion  on possible  uncertainties and  biases  affecting this
    measurement.}),  so  the extinction  although  present, might  not
  account totally  for the late  polarization.  In this  Framework the
  $\theta$ change  of $\sim100^{\circ}$ would just obey  to the chance
  alignment   between  the   asymmetric  expansion   and   the  linear
  polarization induced  by the host  ISM.  Assuming that our  last $P$
  detection  ($t-t_0\sim18.7$ days) is  purely due  the host  ISM, the
  SN~2006aj  intrinsic polarization  at $t-t_0\sim4.5$  days  would be
  higher ($P_{SN}=4.78\pm1.17\%$, see Fig.~\ref{fig3} lower panel).

 An asymmetric distribution of  $^{56}$Ni could alter the SN radiation
 field,  specially  after  the  lightcurve  peak  when  the  $^{56}$Ni
 radioactive decay powers the optical lightcurve.  The collapsar model
 predicts a  subrelativistic disk wind  of neutron and  protons, which
 after cooling would form  $^{56}$Ni (Woosley \& Bloom \cite{Woos06}).
 A  picture to explain  the $\theta$  rotation of  $\sim 100^{\circ}$,
 would  require  an equatorial  deposition  of synthesized  $^{56}$Ni,
 which would alter the $\theta$ measured after the lightcurve peak.

 Another possible  mechanism to  explain the observed  $\theta$ change
 could be  delayed CSM scattered  light. This process was  proposed to
 interpret  the  SN 1987A  polarization  lightcurve  (Wang \&  Wheeler
 \cite{Wang96}).  In  this scenario the polarization  would consist of
 two components, the one due  to photons propagating directly from the
 (possibly aspherical) SN, and the  second one from scattered light by
 a dusty  CSM region.  The epochs  of our observations  would push the
 CSM  scattering region further  than $\sim  2.3$ light-days  from the
 progenitor.   However, this  process should  have originated  a clear
 distortion of  the spectra at  $\lambda \gtrsim 6000$~\AA  ~after the
 lightcurve peak, which is  not observed (Modjaz et al. \cite{Modj06};
 Pian et al. \cite{Pian06}).

\section{Conclusions}
\label{Conclusions}

  We  report  the  detection   of  $R$-band  linear  polarization  for
  XRF~060218/SN~2006aj in observations done since $t-t_0 \sim 3$ to 39
  days. These  observations represent  the first detection  of optical
  polarization in  an XRF.  The  measured high degree  of polarization
  gives  further credence to  the XRF/GRB  off-axis scenario,  since a
  face-on SN would not  likely produce any intrinsic polarization.  As
  Patat et al (\cite{Pata01}) noted  there is a degeneracy between the
  viewing  angle and  the  asymmetry level,  so  a detailed  geometric
  picture based  on only our  data is not  possible.  Our data  show a
  $P\sim4\%$  value  in the  first  $\sim  3-5$  days after  the  GRB,
  followed by a  constant $P \sim 1.4\%$ polarization  phase at $t-t_0
  \gtrsim 13.7$ days, suggesting a  $P$ decay.  $\theta$ shows a $\sim
  100^{\circ}$ rotation at $5.6 <  t-t_0 <13.7$ days.  We propose that
  the $P$ evolution can be  explained by an asymmetric SN.  We discuss
  several ingredients  which separately, or combined,  can explain the
  observed $\theta$ rotation (in  decreasing order of relevance); i) a
  highly  asymmetric high-velocity  SN  ejection, followed  by a  less
  asymmetric SN bulk,  ii) polarization induced by the  host ISM, iii)
  asymmetric distribution  of $^{56}$Ni and iv) scattering  by a dusty
  region placed a few light-days from the progenitor.

\section*{Acknowledgments}
This   study    is   supported   by    Spanish   research   programmes
ESP2002-04124-C03-01 and AYA2004-01515.   JG thanks the hospitality of
the  Donostia  International  Physic  Center  (DIPC).   We  thank  our
anonymous referee for helpful comments.

\end{document}